\documentclass{desyproc}

\begin{document}
\title{The Pressure of Misalignment Axions:
a Difference from WIMPs in Galaxy Formation?}

\author{{\slshape Sacha Davidson$^1$}\\[1ex]
$^1$IPN de Lyon/CNRS, Universit\'e Lyon 1, Villeurbanne, France}

\contribID{Davidson\_Sacha}

\confID{300768}  
\desyproc{DESY-PROC-2014-03}
\acronym{Patras 2014} 
\doi  

\maketitle

\begin{abstract}
Two populations of axions can contribute to cold dark matter: 
the classical field produced via the misalignment mechanism, 
and the modes produced in  the decay of strings. 
The classical field  has extra pressure, 
as compared to WIMPs, which could have observable
consequences  in  non-linear galaxy formation.
\end{abstract}

\section{Introduction}

It is interesting to study whether axion dark matter could
be distinguished from WIMP dark matter, using Large Scale Structure
(LSS) data. It is well known that axion dark matter can be
composed of two components\cite{Sici}: the misalignment
axions and the non-relativistic modes radiated by strings
(here taken to be cold particles). 
Both    redshift like CDM, 
and grow linear density  inhomogeneities  like 
WIMPs. However, as pointed out by Sikivie \cite{SikivieTmn},
the misalignment  axions  have a different pressure from WIMPs, 
which could be relevant during  non-linear stucture formation. 
The consequences of this additional pressure could be reliably
addressed by the numerical galaxy formation community. 
The aim of this proceedings, which is based on \cite{axions2},
is to clarify the relevant variables and equations for
studying non-linear stucture formation with axions.

There has been considerable confusion in the literature about whether
axion dark matter is a  Bose Einstein condensate.  In a
scenario proposed by Sikivie, the    dark matter axions
  ``thermalise'' via   their gravitational
interactions, and therefore form a Bose Einstein
condensate  due to the high occupation number of 
the low momentum modes. Then, Sikivie and collaborators hypothesize
that  a  galactic  halo made of condensate could 
form vortices, which could be observed as caustics in
the dark matter distribution. This interesting scenario,
which proposes an observable signature for axion dark
matter in LSS data, has been studied by many people:
Saikawa and collaborators\cite{japonais} confirmed
in Quantum Field Theory  and General Relativity, the gravitational
interaction  rate  estimated by
Sikivie and collaborators. However, with
Martin Elmer\cite{axions1}, we could not confirm
that the interaction rate was a thermalisation rate
(= generated entropy). Rindler-Daller and Shapiro\cite{RDS}
studied rotating halos of non-relativistic
scalar field, and found that vortices were
energetically favoured for much lighter
bosons than the QCD axion, or for
scalars with repulsive  self-interactions 
(opposite to axions, whose self-interactions
are attractive). 
This proceedings will argue that
the notion of  Bose Einstein
condensation is an unneccessary confusion,
somewhat akin to trying to describe a classical
electromagnetic field  in terms of photons. 
The misalignment axions are a  classical scalar
field,  as such they have a different pressure
from WIMPs, and  this could allow them to
leave distinctive features in LSS data.
First, I review the  formalism and
results of \cite{axions2},  and discuss 
the relation to Bose Einstein condensation after
eqn
(\ref{Tij}).

Structure formation within the
horizon is a classical gravity problem,
so at first sight, a Quantum Field Theory
treatment seems unneccessary.
However, the classical field and classical
particle limits of a QFT are different,
and dark matter axions are composed of both. 
So to obtain a self-consistent 
and well-defined formalism,
where approximations can be catalogued,
and particles and fields can be
included simultaneously,
this proceedings  starts from
the path integral. 
A very beautiful earlier treatment of
axion dark matter in field theory is \cite{NS}. We
will find that it 
is straightforward to describe axion dark matter 
as a  classical field (the misalignment axions)
plus cold axion particles (produced by strings).
The leading order gravitational
interactions of the axion field and particles are simply
found by computing their contributions
to the stress-energy tensor. And while cold axion
particles, in the limit  of negligeable velocity,
 have the stress-energy as WIMPs, the classical
axion field has extra contributions to the
pressure, which could be relevant during non-linear
stucture formation. 

Finally, in section
\ref{sec:evap}, I discuss the rate at which gravity
can move axions between the field and the
cold particle bath.

\section{Formalism}

From a theoretical perspective, it  should be true that
``the path integral knows everything'': our world  
is usually at  a saddle-point of the effective action. This perspective is
rarely useful, because the path integral cannot be solved.
However, I  imagine to follow it here for two reasons: first, it
gives an formal framework which can describe all
aspects of axion behaviour, and second, the axion
is so feebly coupled  that the path integral could
be computed pertubatively,  which organises 
the various corrections to the classical= leading order
solutions.

  I imagine to write  the path integral for the axion field,
and evalulate it in a closed-time-path (CTP)
formalism, as a perturbative expansion in 
Newton's constant $G_N$  and the axion self-interaction
parameter $m^2_a/f^2_{PQ}$. The path integral offers to
decribe axions  via $n$-point
functions. I am particularily interested in
 the one-point function,
which is the classical axion field, and the
two-point function, which in CTP formalism includes
the number  distribution $f$ of axion particles.
Higher point functions can be neglected because
the axion is so feebly coupled. 

Next, equations of motion are required. Since
we are interested in the gravitational interactions
of axions, the leading order (= classical) equations
are  Einstein's Equations, with  contributions
to the 
stress-energy tensor 
from the axion field and particle density. 
These  contributions  will be obtained in
second-quantised Field Theory in flat
space-time, for simplicity.
 The details of the calculation can be
found in \cite{axions2}.
The order $G_N^2$ effects will be discussed 
in section \ref{sec:evap}.

  For  cold axion particles, 
self-interactions  are $\sim \lambda^2 \sim (m_a/f_{PQ})^4$
which is negligeable, so 
$T_{\mu\nu}$ has the form expected for
non-relativistic  non-interacting 
particles
(the same as for WIMPs):
$$
T_{\mu \nu} = \rho U_\mu U_\nu =
\left[
\begin{array}{cccc}
\rho & &-\rho\vec{v}&\\
&&&\\
-\rho\vec{v}&& \rho v^iv^j&\\
&&&
\end{array}
\right] ~~~ ~~~
{\rm where}~~~ ~~~
{\rho}(x) = \int \frac{d^3q}{(2\pi)^3} m_a
f(x,q) ~~~,
$$
which is the expected classical  particle result,
with matter four-velocity  $U^\mu = (1,\vec{v})$.

For the  non-relativistic classical field $\phi = \eta e^{-i(mt+S)}$,
  $T_{\mu \nu}$ is
of the form
\begin{equation}
T_{\mu \nu}  =
\left[
\begin{array}{cccc}
\rho & &-\rho\vec{v}&\\
&&&\\
-\rho\vec{v}&& \rho v^iv^j + \Delta T_{ij}&\\
&&&
\end{array}
\right] ~~~ ~~~
\begin{array}{ccl}
\Delta T_{ij} & = &
 2 \partial_i \eta    \partial_j \eta
  -  \delta_{ij} \nabla \eta  \cdot  \nabla \eta
\\ &&
-  \delta_{ij} \left(
 \rho |\vec{v}|^2 - 2m\eta^2 \partial_t S  
 + \lambda \eta^4 \right)
\end{array}
\label{Tij} 
\end{equation}
where $\rho \simeq 2 m^2 \eta^2$, $v_j\simeq \partial _jS/m$,
and  $\Delta T_{ij}$ is the extra pressure 
intrinsic to the classical scalar field.
There is a part related to the
field gradients,  sometimes refered to
as ``quantum pressure''\cite{PitRev}, and a part due
to self-interactions, which is linear in
the   four-axion 
coupling $\lambda = -m_a^2/(12f_{PQ}^2)$.
Since the self-interaction pressure
of axions is {\it negative},  it
causes axions to clump (like gravity). 
Rindler-Daller and Shapiro \cite{RDS}
argue that this negative pressure discourages
vortices in axion halos.  

Notice that the pressure excess  $\Delta T_{ij}$ 
arises because the misalignment axions are a classical field.
There is no additional requirement of ``bose
einstein condensation''.  In calculations, 
Bose Einstein condensates are described
 at leading order as non-relativistic classical fields\cite{PitRev};
since this  is what the misalignment axions
are, one could also say the misalignment axions
are a bose einstein condensate. However, in
my  opinion,  this comes with baggage of
un-useful \footnote{Statistical mechanics is un-useful because its
about the classical particle limit of QFT, 
and $^4He$ is a poor analogy because axions are
feebly coupled.} intuitions from classical equilibrium
statistical mechanics, and from  well-known strongly interacting
Bose Einstein condensates  such as $^4He$.  So its
simpler and clearer to refer to misalignment axions as
a classical scalar field.

The pressure excess $\Delta T_{ij}$  confirms Sikivie's
expectation that axions could differ from WIMPs during
non-linear structure formation. This can most easily be seen
from the ``continuity'' and ``Euler''  equations obtained from 
$T^{\nu \mu}_{~;\mu}=0$:
\begin{eqnarray}
\partial_t \rho + \nabla\cdot (\rho \vec{v})& =& 0 ~~~ ~~~ 
~~~ ~~~ ~~~~ ~~~ ~~~~ ~~~ ~ ~~~  {\rm continuity}
\nonumber
\\
\rho \partial_t \vec{v}  +
\rho ( \vec{v}\cdot \nabla)  \vec{v}& =& -\rho \nabla \psi 
 -\rho\nabla Q   -\nabla  P_{SI}
~~~ ~~~   {\rm Euler}~~~,
\nonumber
\end{eqnarray}
where $\psi$ is the Newtonian gravitational potential,
 $Q = -\frac{1}{ 2m \eta} \nabla^2 \eta$  describes
the ``quantum kinetic energy''  related to the 
``quantum pressure''  of the classical field, and
$P_{SI} =  2\lambda \eta^4$  is proportional to the pressure arising
from the  Self Interactions of the field. The point is that
the extra pressures generate extra forces on the dark matter
distribution. It would be interesting to numerically simulate 
galaxy formation using fluid equations for the
dark matter, rather than N-body.  This could allow
to identify differences between galaxies made
of  classical  axion  field   versus   cold particles.

\section{Gravitational evaporation of  the field?}
\label{sec:evap}

An important question remains,  before studying galaxy formation
in the the presence of the classical axion field:
can gravity  move axions  between the
field and the particle bath? One could imagine that
during violent moments of galaxy formation, the field gets stirred up,
and evaporates into particles. Or maybe the particles
condense to the field, due to their high occupation
numbers?

This process of ``decay'' of the field into particles
should arise somewhere in the  path integral formalism.
For instance, in the case of a  $\lambda\phi^4$
interaction,  this ``decay'' of the field
 will arise as an imaginary part of loop corrections
to the potential,  in a 2 Particle Irreducible
Closed-Time-Path evaluation of the path integral. 
A more simple-minded estimate was performed
in  \cite{axions2}. 

First, notice that classical  gravity, at ${\cal O}(G_N)$, does not
move the particles between the field and the bath, because
the field and particles contribute independently
to the stress-energy tensor:
$$
T_{\mu\nu} = 
T^{(\phi_c)}_{\mu\nu} + T^{(part)}_{\mu\nu}  ~~~.
$$
Then, one can estimate  the   ${\cal O}(G^2_N)$  cross-section for
axion  scattering  via graviton exchange,  $a a \to aa$. If one
of the incident axions is in the classical field, and both final state
axions are free particles, this would correspond to field ``evaporation.
The cross-section is infra-red divergent, so the choice of infra-red
cut-off is crucial. I claim that a reasonable IR cutoff is
the inverse axion momentum, because on much larger distances,
the graviton will not see  individual axions, but rather
will interact coherently with  a large number of axions. 
Therefore the the two-axion scattering process,
 $a a \to aa$, only occurs  when the momentum  exchange
is of order the axion momentum. In this case, the
rate for the field to evaporate to particles
via gravitational interactions  is found
to be negligeably small, because it is suppressed
by $(m_a/m_{pl})^3$.

\section{Summary}

The axion  misalignment field  has
additional contributions to the pressure, as compared to
a bath of cold particles. This is automatic, no  dynamical process of 
Bose Einstein  condensation is required. 
As suggested by Sikivie, this extra pressure could give
observational signatures  in the structure of galaxies. It would
be interesting to numerically simulate galaxy formation 
with a fluid  code which allowed the  dark matter  to
have pressure, 
to discover whether galaxies made of axion dark matter
look different.


\begin{footnotesize}

\end{footnotesize}


\end{document}